\documentclass[12pt,a4paper]{article}   	

\usepackage[utf8]{inputenc}

\usepackage{graphicx}

\usepackage[UKenglish]{isodate}

\usepackage[a4paper,margin=2cm, marginparwidth=2cm]{geometry}

\usepackage{amsmath,amssymb,amscd}
\numberwithin{equation}{section}

\usepackage{amsthm}
\theoremstyle{definition}

\usepackage{graphicx}
\graphicspath{{./figures/}}

\usepackage[dvipsnames]{xcolor} 

\usepackage[font=small,labelfont=bf]{caption}
\usepackage{float}

\providecommand{\citep}[1]{\cite{#1}}
\providecommand{\citet}[1]{\cite{#1}}

\usepackage[colorlinks,allcolors=blue]{hyperref}

\providecommand{\href}[2]{\texttt{#2}}
\providecommand{\url}[1]{\texttt{#1}}

\newcommand{\bea}{\begin{eqnarray}}
\newcommand{\eea}{\end{eqnarray}}
\newcommand{\beq}{\begin{equation}}
\newcommand{\eeq}{\end{equation}}

\providecommand{\eqref}[1]{(\ref{#1})}

\newcommand{\figref}[1]{Fig.\ \ref{#1}}
\newcommand{\tabref}[1]{Table~\ref{#1}}

\newcommand{\secref}[1]{Section~\ref{#1}}

\newcommand{\appref}[1]{Appendix~\ref{#1}}

\newcommand{\tsenote}[1]{}
\newcommand{\tsecomment}[1]{}

\newcommand{\tsedef}[1]{\textsc{#1}}

\newcommand{\Acal}{\mathcal{A}}
\newcommand{\Ccal}{\mathcal{C}}
\newcommand{\Ecal}{\mathcal{E}}

\newcommand{\Gcal}{\mathcal{G}}
\newcommand{\Kcal}{\mathcal{K}}
\newcommand{\Ncal}{\mathcal{N}}

\newcommand{\Amat}{\mathsf{A}}
\newcommand{\Cmat}{\mathsf{C}}
\newcommand{\Dmat}{\mathsf{D}}

\newcommand{\kav}{\langle k \rangle}
\newcommand{\kbar}{\bar{k}}
\newcommand{\kin}{k^\mathrm{(in)}}
\newcommand{\kintr}{k^\mathrm{(in,TR)}}

\newcommand{\Rin}{R^\mathrm{(in)}}

\newcommand{\sigmabar}{\overline{\sigma}} 
\newcommand{\Deltabar}{\overline{\Delta}} 

\newcommand{\Neff}{N^\mathrm{(eff)}}

\begin{document}

\renewcommand{\thefootnote}{\fnsymbol{footnote}}

%
%
%
%

\begin{center}
 {\Large\textbf{Interdisciplinarity Revealed by }}\\{}
 {\Large\textbf{Transitive Reduction of Citation Networks}}
 \\[\baselineskip]
 {\large 
 H.\ AlMuhanna\textsuperscript{1,2}\footnote{ORCID:  \href{http://orcid.org/0009-0004-2564-0140}{\texttt{0009-0004-2564-0140}} }, 
 \href{https://lbb.ethz.ch/the-group/post-doc/vasiliauskaite--vaiva-dr-}{V.\ Vasiliauskaite}\textsuperscript{1,2,3}\footnote{ORCID:  \href{http://orcid.org/0000-0003-2039-6236}{\texttt{0000-0003-2039-6236}} },
 \href{http://www.imperial.ac.uk/people/t.evans}{T.S.\ Evans}\textsuperscript{1,2}\footnote{ORCID:  \href{http://orcid.org/0000-0003-3501-6486}{\texttt{0000-0003-3501-6486}} }}
 \\[0.5\baselineskip]
 1.\ \href{http://complexity.org.uk/}{Centre for Complexity Science}, Imperial College London, SW7 2AZ, U.K.
 \\
 2.\ \href{http://www3.imperial.ac.uk/theoreticalphysics}{Abdus Salam Centre for Theoretical Physics},  Imperial College London, SW7 2AZ, U.K.
 \\
 3.\ Laboratory of Biosensors and Bioengineering, Institute for Biomedical Engineering, Gloriastrasse 37/39, 8092 Zurich, Switzerland
 \\[0.5\baselineskip]
 29\textsuperscript{th} July 2025
\end{center}

\begin{abstract}
We investigate the impact of transitive reduction on citation networks. Our hypothesis is that documents which lose fewer citations under transitive reduction are likely to be interdisciplinary, while a large loss of citations suggests a document is primarily cited within a single discipline. We test this hypothesis by using an artificial model of a citation network and by using data on citations from three sources: academic papers, court decisions and patents. 
Where needed, we applied modularity-based clustering techniques on a network defined using bibliographic coupling to classify documents by topic. A cluster-dependent measure was then used to classify the nodes as interdisciplinary or intradisciplinary. Our results provide strong support for our hypothesis in three of the four cases, with somewhat weaker but still positive support in the case of patents.
\end{abstract}

\setcounter{footnote}{0}
\renewcommand{\thefootnote}{\arabic{footnote}}


\section{Introduction}

Interdisciplinarity plays a crucial role in the advancement of knowledge, as many influential ideas emerge at the intersection of distinct research fields \cite{P63,GLNSST94,PR09,O19,RWL23,WCCC23,YWL25}. When analysing documents across multiple disciplines, such as different research fields, we expect documents within the same discipline to cluster together. However, some documents have an impact across multiple disciplines, while not strictly falling within a single field. A notable example is the AlphaFold paper by DeepMind \cite{JEPGF21}, which uses transformer-based neural networks to predict a protein's 3D structure from its amino acid sequence. This landmark study combines advances in artificial intelligence and molecular biology, and the work itself is not easily classified within a single discipline.

A commonly used measure to assess the influence of a document is its citation count. Documents with high citation counts are often considered significant, especially when it comes to academic papers. However, this alone may not fully capture interdisciplinarity \cite{CLZ23}. 
Citations may arise due to various motivations, including social or habitual factors, rather than genuine intellectual influence \cite{B86,MM89}. In addition, many citations may be redundant since authors frequently draw their information directly from more recent up-to-date documents, even if they cite the older sources \cite{SR03,SR05}.

However, the influence of a document is not simply defined by its citations. Rather, a document has a wider, often indirect, influence on how a subject evolves. These aspects are well captured by a network \cite{GST64,P65b,ZS15}, and in this context, citation networks offer the best way to represent these features. In citation networks, nodes represent documents, such as academic papers, while edges denote citations. Here we adopt a convention where edges point backward in time from newer documents to the older ones they cite. Consequently, the in-degree of a node represents its citation count, while the out-degree reflects the number of references in its bibliography. 
The time constraint on the edges means that a citation network is a special type of network known as a directed acyclic graph (DAG), whose unique properties we will exploit. More broadly, citation networks provide a valuable framework for analysing the structural flow of information within research \cite{GST64,P65b,ZS15,G19,SWTE20,HPEO24,ZGWWZ25}.

\begin{figure}[htb] 
	\centering
	\includegraphics[width=\linewidth]{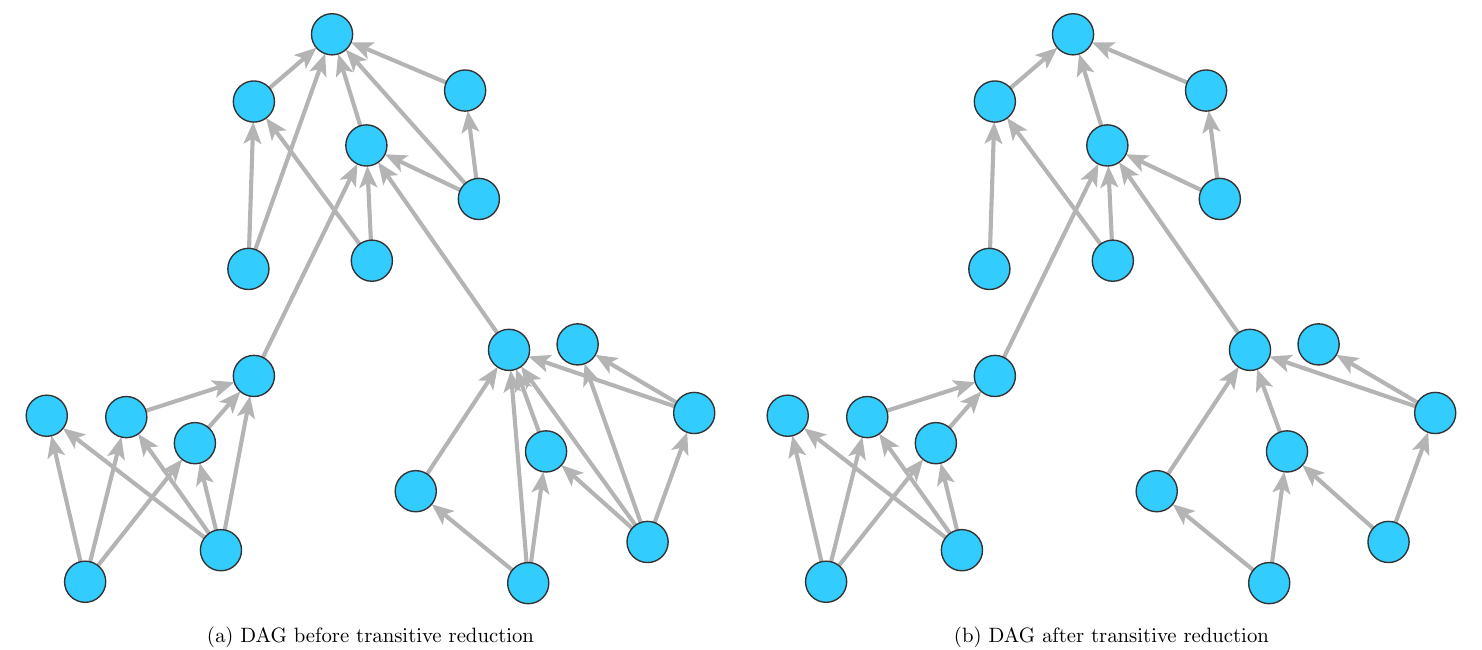}
	\caption{Example of performing transitive reduction on a Directed Acyclic Graph. Redundant edges are removed but the overall connectivity of the network is preserved.} 
	\label{tr}
\end{figure}

Our focus is on the idea that many citations from a given document to older sources are redundant and do not necessarily imply a direct flow of information. In terms of the citation network, there are edges that can be removed in a process known as \tsedef{transitive reduction}, illustrated in \figref{tr}, without changing the connectivity of nodes. Transitive reduction gives the minimal structure needed to preserve the flow of information. 
One of the special properties of a directed acyclic graph, such as a citation network, is that there is no uncertainty in the transitive reduction process; edges are either essential to maintain connectivity or are redundant and can be removed. This is a distinctive feature for DAGs, as opposed to other directed graphs, where cycles can complicate the transitive reduction process.

Our proposal is that documents exhibiting minimal changes in their citation count after transitive reduction tend to be multidisciplinary. The change in bibliography could also be studied; however, we suspect that the citations a paper receives are more informative than the works it cites. 
For example, the impact of a foundational paper could only become apparent through the emergence of subfields that develop in response to it. 
We test this hypothesis using citation datasets and an artificial model, employing richness, an information-theoretic measure of diversity, to quantify interdisciplinarity. If our proposition holds, nodes with high diversity values should exhibit low citation loss post-transitive reduction.

\section{Datasets and Models}\label{s:datamodel}

To validate our approach, we apply our methodology to citation networks generated by an artificial model: the Preferential Directed Acyclic Graph (PDAG) model. This is a generalisation of previous models and numerical implementations \cite{P76,N10,GAE15a,ECV20}.  
The nodes $n_i \in \Ncal$ represent documents and are labelled by integers $i\in\{0,1,2,\dots,N-1\}$, where the label of the $i$-th node is also the time of publication of a document represented by node $i$. 
Each document $n_i$ is assigned uniformly at random to one field (cluster), denoted by the label $c_i \in \Ccal$ drawn from $\Ccal$, the set of labels of the different fields. 
The directed edges $(n_i,n_j)$ from node $n_i$ to node $n_j$ represent links from the bibliography of the newer document $n_i$ to the older document $n_j$. As a result, for any directed edge $(n_i,n_j)$, we have that $i>j$, ensuring an acyclic structure where the node label $i$ also serves as a time coordinate for node $n_i$. This defines a total order on this directed acyclic graph.

Nodes are also assigned to be one of two types; with probability $p_\mathrm{mf}$ nodes are a multi-field paper, otherwise they are a single-field paper. 
Single-field papers in field (cluster) $f$ can only cite single-field papers within their own field $f$. However, part of their bibliography will consist of multi-field papers $n_j$ from field $c_j$, provided document $n_j$ has cited at least one paper in field $f$. Multi-field papers can cite papers from any field. 
The entries $n_j$ in the bibliography of any paper $n_i$ are chosen with a probability roughly proportional to $(\mu+\kin_j)$ where $\kin$ in the in-degree and $\mu =12$ is a model parameter. This gives a fat-tailed citation distribution as can be seen from simple generalisations of the Price model \citep{P76,N10,GAE15a}. In addition, a fraction $p=0.45$ of the bibliography comes from recent papers, here coming from the most recent $2200$ or so papers of each field, i.e.\ when citing a recent paper, we have that $2200 \gtrsim (i-j) >0$. The remaining 55\% of papers come from papers of any age. This is to simulate the well-known feature \citep{GAE15a,ECV20} that recent papers appear more often in bibliographies than found in simple models using attachment probabilities of the form $(\mu+\kin_j)$. 
The parameter values used were chosen to produce a citation network with characteristics similar to the hep-th arXiv data set discussed below. 
A more detailed description of the model is given in \appref{as:MFCNmodel}.

We use three datasets to give us the citation network of three different types of documents. The first dataset gives us the citation relationships between papers posted on the High Energy Physics Theory (hep-th) category on the arXiv preprint repository between 1992 and 2003. This dataset, sourced from the Cornell KDD Cup website \cite{GGK03,KDDcup}, contains 27,770 nodes representing academic papers and 352,807 edges representing citations.

The second dataset represents the legal citations found in majority opinions of the United States Supreme Court between the years 1754 and 2002. The dataset, collected and developed by Fowler and Jeon as described in \cite{FJ08, FJSJW07}, consists of 25,417 nodes representing opinions and 216,738 edges representing citations. 

The last dataset gives us a citation network of patents. For this we use data from the National Bureau of Economic Research (NBER) \cite{HJT01}, consisting of patents registered in the U.S.\ between 1975 and 1999, with 3,774,768 nodes representing granted U.S.\ patents and 16,518,949 edges representing citations. This also has a single subject category assigned to each patent.

The advantage of using these three datasets is that the motivation for adding a reference to a document is not necessarily the same in these three cases, giving us a broader test of our hypothesis and methods. For academic papers represented by the hep-th dataset, there is little formal requirement for a rigorous standard in the choice of quoted papers. So while references to key earlier developments are often included, it is also clear that citations frequently reflect other pressures, such as personal relationships or popularity, as part of a so-called ``citation culture'' \cite{W99}.
On the other hand, legal citations in the United States Supreme Court follow much stricter guidelines; references are made based on the legal principle of ``precedent'', which restricts citations to relevant court rulings, maintaining consistency within the justice system \cite{CSJW10}. Finally, patents follow a different citation logic. The main aim of citations in patents is to prove an invention's novelty, and there is a legal requirement to cite ``prior art''; existing patents relevant to the current one. Still, many references in a patent can be to highly influential patents from companies such as Google or IBM \cite{PRA14}. Additionally, self-citation is a common phenomena among large organisations; it broadens the inventor's patent portfolio and acts as a sort of protection from legal issues by ensuring the robustness of their intellectual property \cite{BCM09}. 

Note that some of our datasets have `future-pointing' edges. These can arise because several publishing dates can be linked to a single document \cite{HBC15}, or due to errors in the recorded dates. In such cases, we reverse the direction of the affected edges. Citations from a document to itself are rare (less than 0.15\% of edges), but they do occur in the datasets, and we remove these edges. We then extract the largest weakly connected component \cite{C21} from the resulting citation network. For the NBER patent dataset, the network was considerably larger than the other examples, so we further reduced this dataset by removing nodes with a degree less than ten. Further information on the source of the data and the input files used in our analysis is given in \appref{as:datasets}.

\section{Methods} \label{s:methods}

First, we need to identify which documents are multidisciplinary and which are not. To do this, we need to identify the field or topic associated with each document. Then, we need to define measures that show us if the change in in-degree after transitive reduction is smaller for interdisciplinary documents than it is for documents working within a single field.

\subsection{Cluster labels}\label{s:clusters}

An interdisciplinary document will have influence across multiple topics. Before we can even quantify interdisciplinarity, we must first define what constitutes a topic, a research field for academic papers and patents, and an area of law for USSC judgements. We will define the topics/fields by partitioning the set of documents, i.e.\ to each node $n_i$ we assign a label $c_i$ where the number of distinct labels is $N_c$.

One option would be to use any subject labels provided in the metadata for the documents. 
For the patent data, we have such labels, and due to the size of this dataset, we use these labels to define our cluster set.
However, for the hep-th and USSC datasets, the documents are not assigned to a specific category. Therefore, we have a bottom-up strategy, creating our own field/topic label for each document based on the network structure. This can be applied to all datasets to ensure that we can make a fairer comparison across our citation networks.

In the case of our artificial model, every paper is already assigned to be either a multi-field or a single-field paper, as well as having an explicit field or topic assignment. However, the stochastic nature of the model means that there is a chance that, in particular instances, a node may not behave exactly as the predefined label suggests. So we choose not to use the intrinsic node labels for data produced from the model. Instead, we use the same bottom-up technique as used on the hep-th and USSC datasets.

In network analysis, clustering of nodes is known as ``Community Detection''. In the context of a network, a community or cluster is assumed to have a much larger number of edges between nodes in the same cluster than there are edges between nodes from different clusters \cite{C21}. Various community detection methods exist, but we employ the modularity-based Louvain method \cite{BGLL08}, 
which we use to cluster nodes based on the similarity of their citation patterns \cite{BK10}.

In this method, we construct a new undirected network using bibliographic coupling \cite{K63,K63a,BK10}. Let the adjacency matrix for the citation network be the matrix $\Cmat$ where $C_{ij}=1$ if there is an edge from node $n_j$ to $n_i$, i.e.\ document $n_j$ lists the older document $n_i$  in its bibliography. The new network retains the same nodes as the original DAG, so we will denote the weight of an edge between $n_i$ and $n_j$ by the entry $A_{ij}$ in the adjacency matrix $\Amat$ of the bibliographic-coupling network. 
Each pair of nodes $n_i$ and $n_j$ is connected by an undirected but weighted edge if have at least one document in common in their bibliographies. 
Suppose node $n_k$ is cited by both $n_i$ and $n_j$, so that $C_{ik}=C_{jk}=1$. 
This node $n_k$ will contribute the inverse of its in-degree in the original citation network, $1/\kin_k$, to the weight of the bibliographic-coupling link $A_{ij}$. This avoids undue contributions from highly-cited documents. So we have that 
\begin{equation}
	A_{ij} = \sum_k \frac{1}{\kin_k}C_{ki}C_{kj}
	 \, ,
	 \quad
	 \kin_k = \sum_i C_{ki} \, .
\end{equation}

Finally, we apply the Louvain community detection algorithm \cite{BGLL08} (any other community detection method could also be used). The algorithm finds a partition of the nodes into communities by finding an approximate maximum for modularity $Q$ \cite{NG04,BGLL08,C21}, which measures how well the network is partitioned into clusters. Modularity $Q$ is defined as \begin{equation}
    Q = \frac{1}{2W} \sum_{i,j} \left( A_{ij} - \frac{s_i s_j}{2W} \right) \delta(c_i, c_j) 
    \, , \quad
    s_i = \sum_j A_{ij}
    \, , \quad
	W = \frac{1}{2} \sum_i s_i \, .
\end{equation}
Here, $\delta(c_i, c_j)$ is one if nodes $i$ and $j$ belong to the same cluster ($c_i=c_j$), and zero otherwise.

\subsection{Interdisciplinarity measures}

Once we have a unique cluster/field/topic label $c_i$ for each document $n_i$, we can quantify its interdisciplinarity. To do this, we look at the distribution of the cluster labels of the neighbours of node $n_i$ in the citation network, using measures of the diversity of their fields \cite{J06, GVDPER23}. 

The simplest diversity measure we use is \tsedef{richness}, which is simply the number of different cluster labels associated with a document $n_i$. In this case, we will use the discipline or cluster label $c_j$ associated with each of the neighbours $n_j$ of a node $n_i$ in the original citation network. 
Since our focus is primarily on the in-degree, it makes sense to use the in-richness $\Rin_i$ of a node $n_i$ as our diversity measure, where 
\beq
 \Rin_i  = \big| \{ c_j | (n_j,n_i) \in \Ecal  \} \big| 
 \, ,
 \label{e:Rdef}
\eeq
where $(n_j,n_i)$ is an edge from node $n_j$ to $n_i$ in the set of edges $\Ecal$ of the citation network.

\begin{figure}[H]
    \centering
    \includegraphics[width=0.5\linewidth]{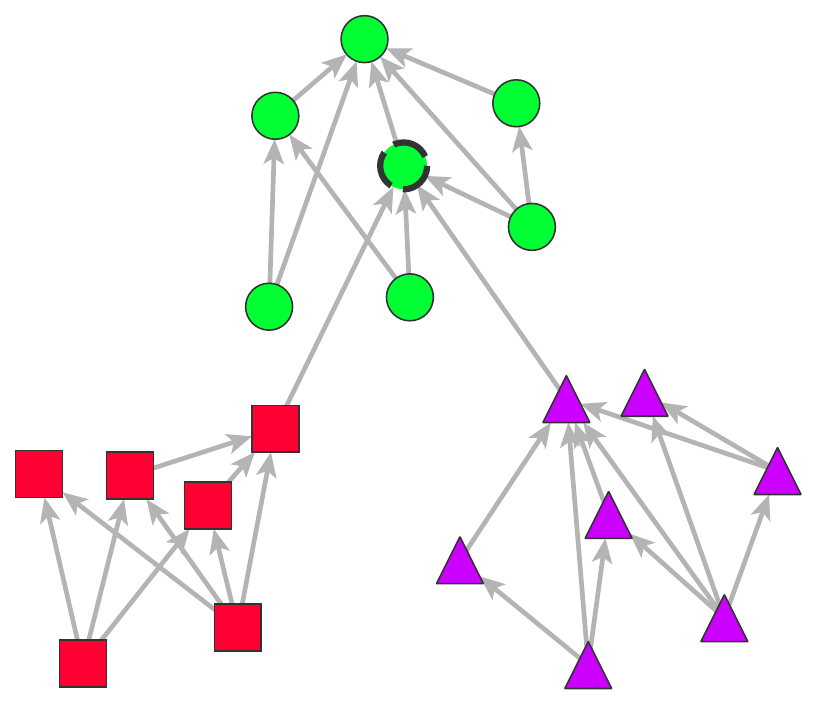}
    \caption{An example of of diversity measures. The node $n_i$ outlined in black has an in-Richness of $\Rin_i=3$, as it is cited by $3$ distinct clusters.}
    \label{fig:enter-label}
\end{figure}

We then need to identify the interdisciplinary documents in our data. Based on the diversity measure for each document, we split the documents into three classes: single-field (SF) documents, multi-field documents (MF) or an intermediate class (IF). This last class is needed because multidisciplinarity is not a simple binary concept in practice, and most documents should really lie on a spectrum between pure single-field documents and maximally interdisciplinary documents.

In our case, we will use the in-richness $\Rin_i$ of each node $n_i$ to determine single-field and multi-field documents. 
It is not so clear where to draw the boundary in richness values between single- and multi-field documents, reflecting the reality that some documents may just be ``mostly'' single-field documents or ``slightly'' interdisciplinary. So we set the criteria that \emph{only} documents with in-richness of one will be considered single-field documents, i.e.\ if $\Rin_i=1$, we will say that document $n_i$ is a single-field document. While only documents with in-richness of three or more $\Rin_i\geq3$, will be considered to be multi-field documents. This means we will not study documents with richness of two as we assume their true nature is unclear from our measurements. 

Our artificial model provides a good example of our motivation for these boundaries. In principle, in the PDAG model, each single field document only cites documents in the same field. That is true for citations from single-field to single-field documents. However, a multi-field document $n_j$, even though it is assigned in the PDAG model to one field given by the value $c_j$, behaves in the model as if it is a member of many fields.  In particular, a multi-field document $n_j$ can cite and be cited by a single-field document $n_i$ regardless of the value of $c_i$. This means a single-field document might be cited by just one multi-field document from a different field, giving that single-field document a richness of two. In the PDAG model, this occurs reasonably often, and this mimics what might happen with real data. Authors may not consider their document to be multidisciplinary, but later authors from a wide range of fields may find value in it.

The network-based approach using modularity can leave us with a large number of clusters, though most contain a small number of documents. Since we can control the number of clusters produced, we need to estimate the number of clusters with a significant number of documents, discounting those with negligible document counts. 
This is because these relatively large clusters dominate our interdisciplinarity statistics, while smaller clusters do not contribute much. 
One measure we use is the h-index $h$ defined over the cluster size distribution $\{n_f\}$, where $n_f$ is the number of documents in the cluster (field) labelled $f$.  
The h-index $h$ is the largest number of clusters which contain at least $h$ documents \cite{H05}. Formally, if $N(\nu)$ is the number of clusters (fields) that contain at least  $\nu$ documents, then
\beq
 h = \max \big\{ \nu \in \mathbb{N} \mid  N(\nu) \geq \nu  \big\} \, , \qquad
 N(\nu) := \big| \{  f  | n_f  \geq \nu \, , f  \in \Ccal \} \big| \, .
\label{e:hdef}
\eeq 
Here $\Ccal$ is the set of labels of the different clusters. 


In addition, we employ Shannon diversity $\Neff$ as our second estimate of the effective number of clusters. 
We use the fraction of nodes $n_f /N$, where $N = \sum_f n_f$ is the total number of nodes in a given cluster $f$, to define the Shannon entropy for a given set of clusters. 
The Shannon diversity $\Neff$ is then given by the exponential of the Shannon entropy, where
\beq
  \Neff = \exp \left( 
  -\sum_{f \in\Ccal} \frac{n_f}{N} \ln \Big(\frac{n_f}{N} \Big)
  \right) 
  \, .
 \label{e:Ddef}
\eeq

\subsection{Analysis of degree change }

Our proposition states that a minimal change in the in-degree (or citation count) of a node after transitive reduction is an indicator of its interdisciplinarity. Since citation counts vary widely across papers, we bin nodes by their original in-degree to control for degree-dependent effects. Bin $b$ is defined to contain all nodes with in-degree between $(k_{b-1}+1)$ and $k_b$. For display purposes, results for each bin are shown at a location given by the degree $\kbar_b$ corresponding to the midpoint of the bin, defined as
\begin{equation}
 \kbar_b  = (k_{b-1}+k_b)/2
 \, .
 \label{e:kbar}
\end{equation} 
Formally, we define 
\begin{equation} \label{e:Bb}
    B_b = \{ n \mid n \in \Ncal , \;\; k_{b-1} < k^{\text{(in)}}_n \leq k_{b} \} 
\end{equation}
where $\Ncal$ is the set of nodes and $k^{\text{(in)}}_n$ is the in-degree of node $n \in \Ncal$. We define $k_b$ to be the smallest degree such that the bin contains at least $n_{\min}$ documents, choosing $n_{\min}=100$ in all cases except for the NBER patents, where we use $n_{\min}=500$.
That is
\begin{equation} 
    k_b = \min\Kcal_b
    \qquad \text{where} \quad
    \Kcal_b = \Big\{ \kappa \Big| \sum_{k=k_{b-1}+1}^{\kappa} n(k^{\text{(in)}}) \geq n_{\min} \Big\} \, .
    \label{e:kb}
\end{equation}
We are comparing the difference in the behaviour of single-field documents, which are represented by the set of nodes in $\mathcal{N}^{(SF)}$, with multi-field documents, represented by the set\footnote{Note that documents cannot be of both types so $\mathcal{N}^{(SF)} \cap \mathcal{N}^{(MF)}=\emptyset$ but we also allow for a situation where not all documents are classified in this way, so $\mathcal{N}^{(SF)} \cup \mathcal{N}^{(MF)}$ may not contain all the nodes.} of nodes in $\mathcal{N}^{(MF)}$. 
To make our comparison, we split each bin into two subsets, $B_b^{(t)}$, where
\begin{equation}
	B_b^{(t)} = B_b \cap \mathcal{N}^{(t)} 
	\, ,
	\quad 
	t \in \{ \mathrm{SF}, \mathrm{MF} \}
	\, .
	\label{e:Bbt}
\end{equation}


We find the in-degree of a node $n$ before and after transitive reduction, $\kin_n$ and $\kintr_n$ respectively. We then measure  the change $\Delta_n$ in the in-degree of a node $n$ before and after transitive reduction, $\Delta_n = \kin_n - \kintr_n$. 
For a given bin $b$ and a given subset of papers of type $t$, we estimate the mean in-degree change $\overline{\Delta}_b^{(t)}$ and associated standard deviation $\sigma_b^{(t)}$ from the average in-degree change $\Delta_n$ of nodes of type $t$ within bin $b$, that is
\begin{eqnarray}
    \Deltabar_b^{(t)} 
    &=& 
    \frac{1}{|B_b^{(t)}|} \sum_{n \in B_b^{(t)}} \Delta_n^{(t)} \, ,
    \label{e:Dbt}
    \\
    \big( \sigma_b^{(t)} \big)^2
    &=&
     \frac{1}{|B_b^{(t)}|-1} \sum_{n \in B_b^{(t)}} \big( \Delta_n^{(t)} - \Deltabar_b^{(t)} \big)^2 \, .
\end{eqnarray}

Finally, to compare the effect of transitive reduction on single-field (SF) and multi-field (MF) nodes, we compute the z-score $z_b$ of the in-degree change per bin $b$
\begin{equation}
	z_b =
	\frac{ \overline{\Delta}_b^{(SF)} - \overline{\Delta}_b^{(MF)} }
	{ \sqrt{ (\overline{\sigma}_b^{(SF)})^2 + (\overline{\sigma}_b^{(MF)})^2 } }
	\label{e:zbdef}
\end{equation}
where $\sigmabar_b^{(t)}= \sigma_b^{(t)} / |B_b^{(t)}|^{1/2}$ is the estimate of the standard error of the mean for nodes in bin $b$ of type $t$.

This z-score quantifies how many standard deviations the citation loss of multidisciplinary nodes differs from that of single-field nodes, helping to assess whether interdisciplinarity correlates with lower citation loss after transitive reduction. A z-score of order one or more for each bin supports our hypothesis.

\section{Results}

We choose not to consider nodes with in-degree $\kin<10$ when assessing changes in the in-degree after transitive reduction. Nodes with low in-degree are very limited in the way they can change their few citations under transitive reduction, and so these documents have very noisy statistics. A similar approach to citation network analysis has been adopted elsewhere, for example, see \cite{CR09}. Some of the plots here display results only up to a certain bin size as there are no intra-disciplinary nodes for higher bin values.

\begin{figure}[H]
	\includegraphics[width=\linewidth]{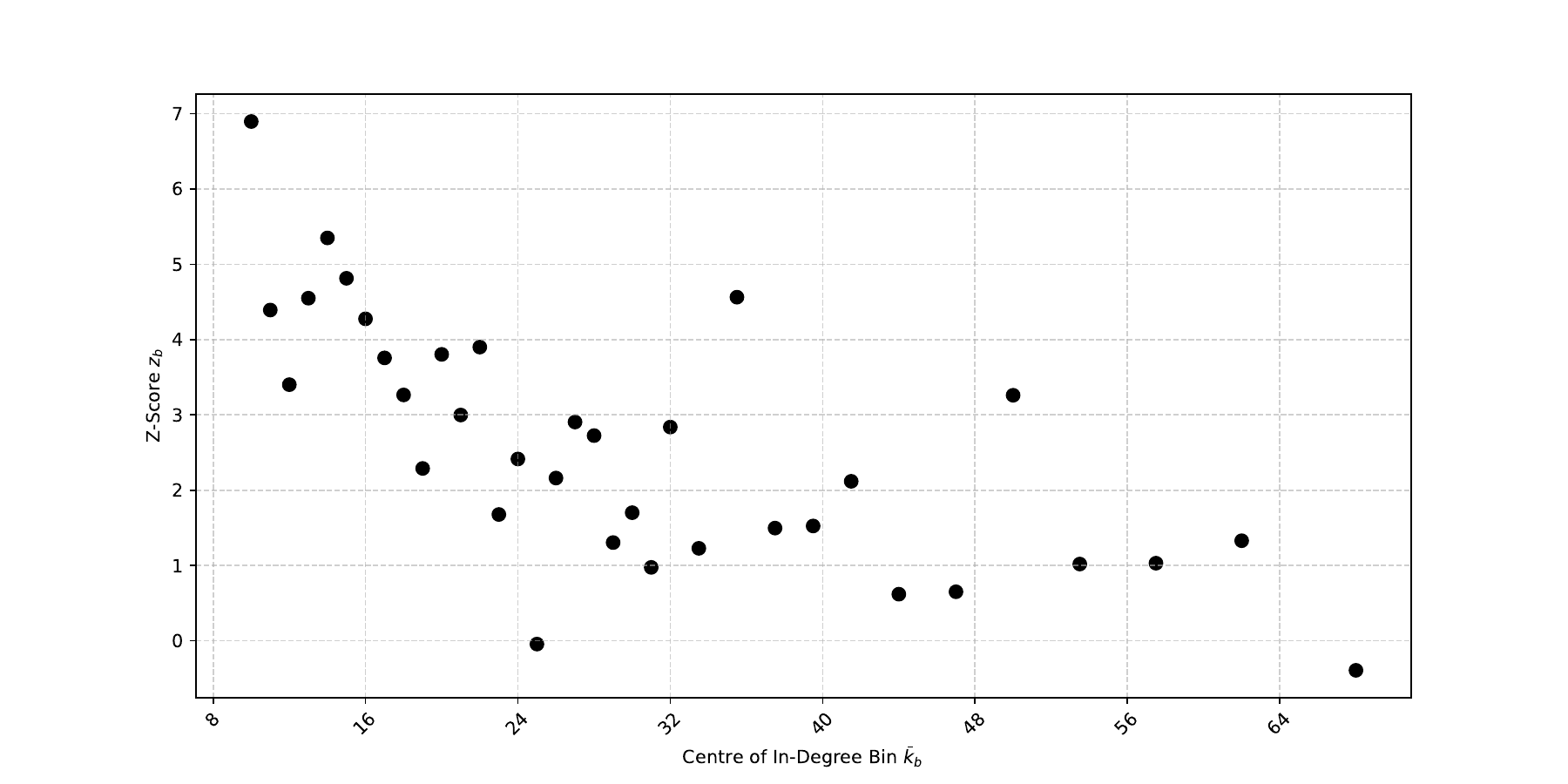}
	\caption{A plot of z-score $z_b$ \eqref{e:zbdef} against bin centre $\kbar_b$ \eqref{e:kbar}  for the PDAG model with $N=27000$ nodes and average degree $\kav=11.95$. Using clustering on the network of bibliographic coupling, we found $N_c = |\Ccal|=130$ clusters with an h-index of $h=10$ and an effective number of clusters $\Neff =10.4$. 
		}
	\label{f:zscoretim}
\end{figure}

The results from our artificial PDAG model, shown in \figref{f:zscoretim}, indicate that all but two of the bins have positive z-scores significantly greater than zero. This suggests that in our PDAG model, multi-field nodes lose fewer citations after transitive reduction. 
However, there is a clear trend toward lower z-scores for bins containing higher-degree documents. These tend to have a smaller number of documents in each bin, but also a wider range of in-degrees for each document, making these statistics inherently noisier than the lower degree bins.

\begin{figure}[H]
	\centering
	\includegraphics[width=0.9\linewidth]{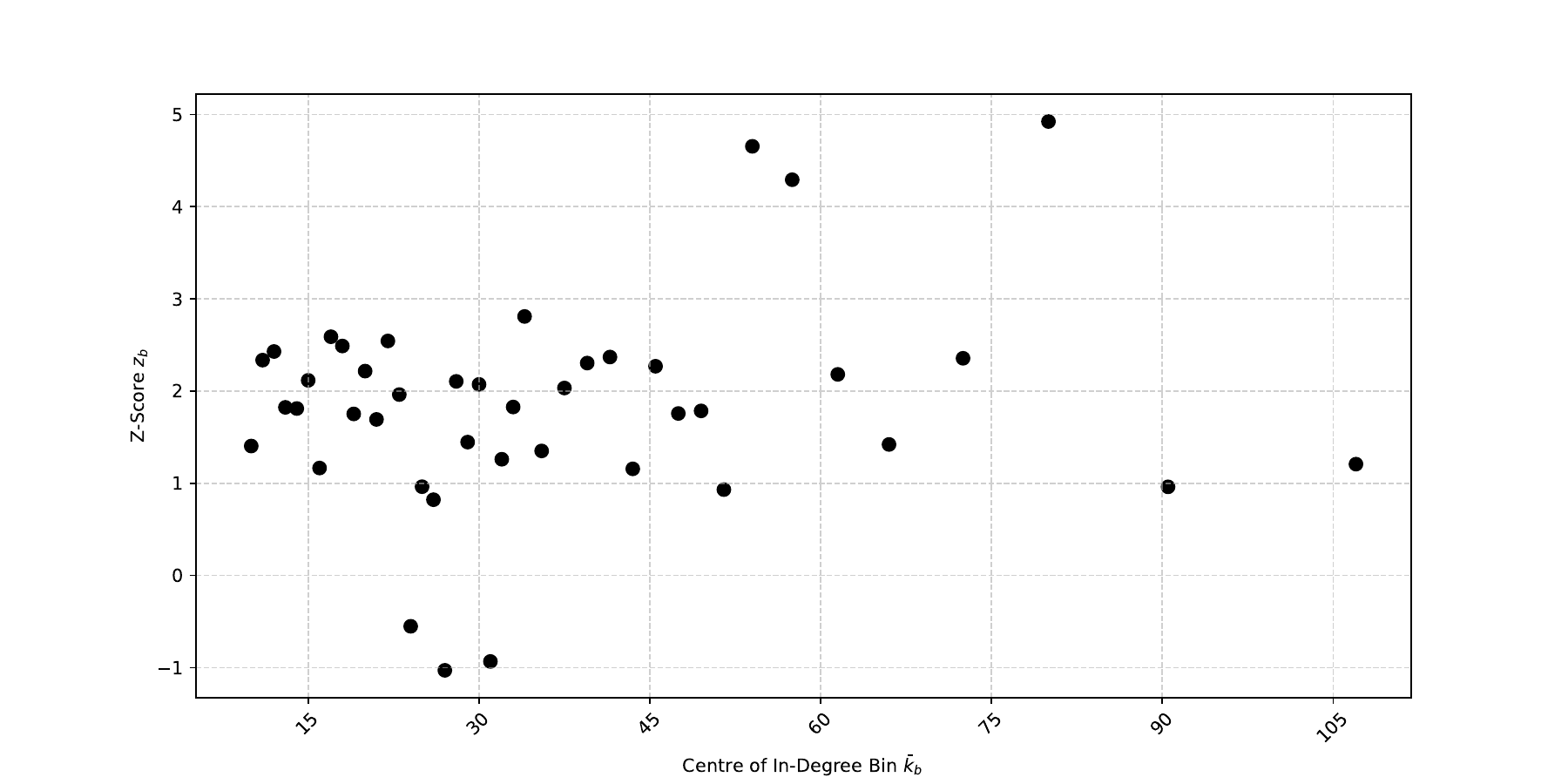}
	\caption{A plot of z-score $z_b$ \eqref{e:zbdef} against bin centre $\kbar_b$ \eqref{e:kbar}  for the hep-th DAG using modularity-based clustering on a bibliographic-coupling network, with $N_c=|\Ccal|=3091$. The h-index is $h=28$ and the effective number of clusters is $\Neff=48.8$.
	}
	\label{f:zscorehep-thlouvain}
\end{figure}

We now turn to analyse our first citation network based on real data, the hep-th papers from arXiv. 
The resulting z-score plot is shown in \figref{f:zscorehep-thlouvain}. All but three of the forty-two bins have positive z-scores, and all but one of those have a z-score of at least $1.0$. Again, this is strong evidence that multidisiplinary papers in hep-th lose significantly fewer citations under transitive reduction than single-field papers. 

By way of comparison, this network-based approach to clustering is significantly more informative than the same analysis based on text-based clustering of hep-th papers using their abstracts, as discussed in \appref{as:textclustering}. We suspect that the hep-th abstracts are too short to properly distinguish the topics of hep-th papers, making the accurate identification of multidisiplinary papers difficult.

\begin{figure}[H]
	\centering
	\includegraphics[width=0.9\linewidth]{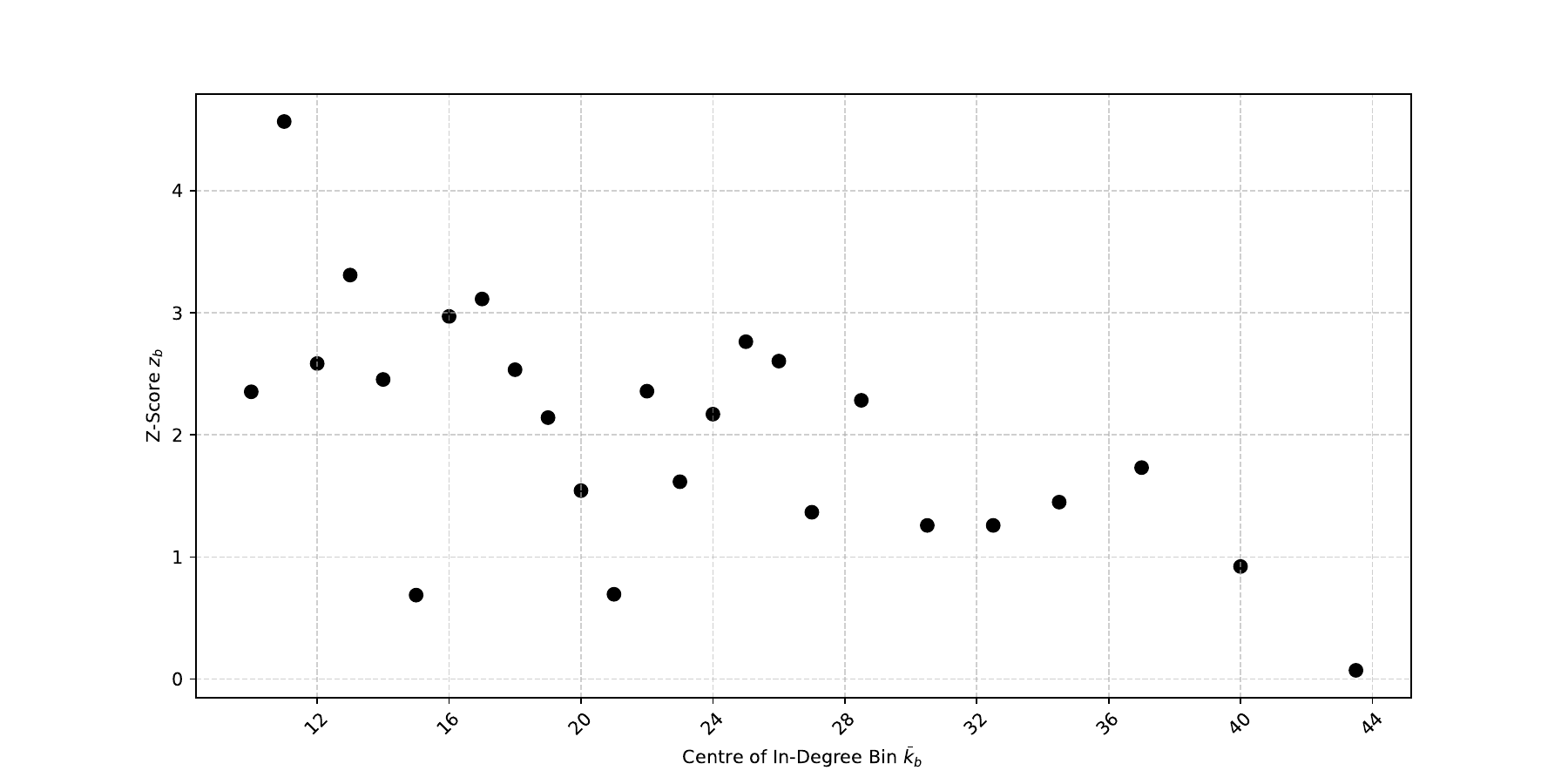}
	\caption{A plot of z-score $z_b$ \eqref{e:zbdef} against bin centre $\kbar_b$ \eqref{e:kbar}  for the USSC DAG using modularity-based clustering on a bibliographic-coupling network with $N_c=6338$. The majority of z-scores are positive but seem to decrease for larger bin centres. The h-index is $h=19$ and the effective number of clusters is $\Neff=103.8$.}
	\label{f:zscoreUSSC}
\end{figure}

All the z-scores observed for the court documents in the USSC DAG in \figref{f:zscoreUSSC} are positive. Further, all but three of the twenty-five bins have z-scores of $1.0$ or more, so the positive values are statistically significant. This gives strong support for our hypothesis in the context of these legal opinions.

\begin{figure}[H]
	\centering
	\includegraphics[width=0.9\linewidth]{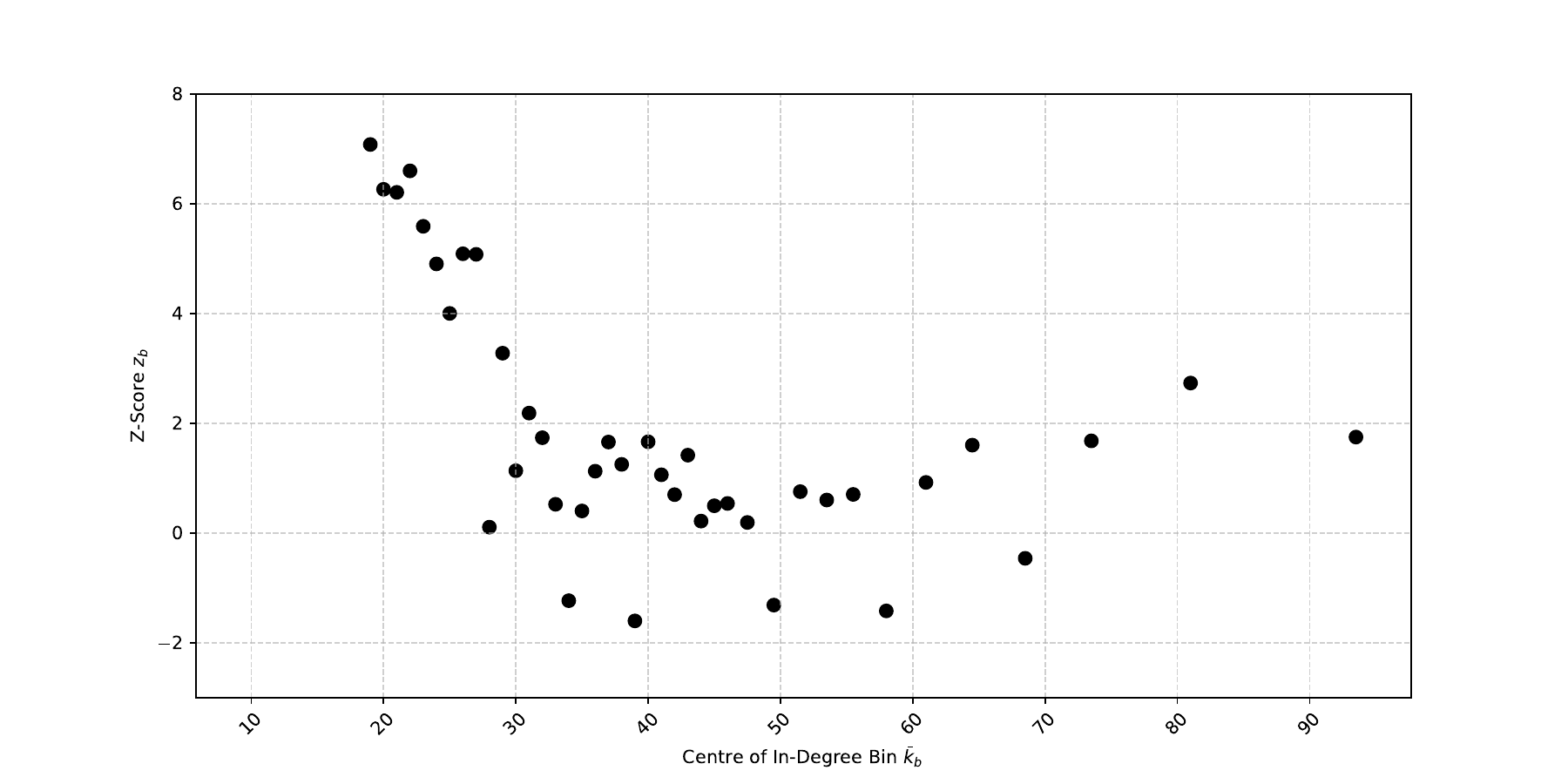}
	\caption{A plot of z-score $z_b$ \eqref{e:zbdef} against bin centre $\kbar_b$ \eqref{e:kbar}  for the NBER DAG. The dataset included a pre-assigned subcategory for each patent, which we treated as a cluster, giving $N_c=36$. Note that the bins with degree $\kbar_b$ between $10.0$ and $18.0$ inclusive have z-scores above $8.0$ and are not shown here for clarity.}
	\label{f:zscoreNBER}
\end{figure}

Finally, for the patents in the NBER DAG, the results in \figref{f:zscoreNBER} are less persuasive. The biggest problem is that the change in in-degree before and after transitive reduction, $\Delta_n$, is generally much lower for patents than academic papers or court judgements \cite{CGLE14} (also see \appref{as:datasets}). This makes any dependence of this change $\Delta_n$ on the multidisciplinary nature of a document harder to see. Yet, it is still true that all but five of the forty bins shown have a positive z-score. However, there is a larger number of bins with z-scores between zero and one than can be seen in previous examples, and more of these values are not significantly different from zero. There is also a strong dependence on the degree of the bin $\kbar_b$. As before, the bins with small degrees, where all documents also have the same in-degree and where there are larger numbers of documents in each bin, exhibit very high z-scores. For higher-degree bins, which tend to have the minimum number of papers in each bin and a wider range of in-degree values for each document, we observe that the z-scores are noisier and show a weaker effect compared to the previous datasets.
Nevertheless, the vast majority of bins display at least a weak positive z-score, with many showing a strong effect. So even with patents, there is good support for the hypothesis that multidisciplinary patents lose fewer citations under transitive reduction than single topic patents.

To provide a simple summary of these results, we show in \tabref{f:zscore table} the average of the z-scores for all the bins shown for each dataset.  
These average z-scores, while giving more weight to highly cited papers, show that there is indeed a significant difference between the in-degree change for multi-field nodes in comparison to their single-field counterpart. The range of average z-scores across DAGs could be ascribed to attributes inherent to the dataset, such as the different citation styles for different types of documents, size of the dataset and effects of the clustering method. We observe that the mean of the interdisciplinary difference is at least 1.6 standard deviations above the mean of the intradisciplinary difference across the bins, again validating our proposition. 

\begin{table}[H]
	\centering
	\begin{tabular}{c|c}
		\textbf{Citation Network} & \textbf{Median z-Score} $\tilde{z}$ \\ \hline
		PDAG    & 2.4     \\ 
		hep-th text clustering (\appref{as:textclustering})    & 1.0     \\ 
		hep-th modularity clustering   & 1.8     \\ 
		USSC    & 2.2     \\ 
		NBER    & 1.7     \\ 
	\end{tabular}
	\caption{The median z-score for every DAG. 
    We only include bins with $\kin \geq 10$ and where the z-score is well defined (some very high degree bins have no multidisciplinary papers).
	}
	\label{f:zscore table}
\end{table}

\section{Discussion and Conclusions}

We started from the observation that there is often a substantial decrease in the citation count of documents after transitive reduction \cite{CGLE14, V20},  
something confirmed in the statistics here (see \appref{as:datasets}). 
However, the loss of citations after transitive reduction is very uneven; some documents can lose a very large fraction of their incoming edges, some very few.  

\begin{figure}[htb] 
	\centering
	\includegraphics[height=0.75\textwidth]{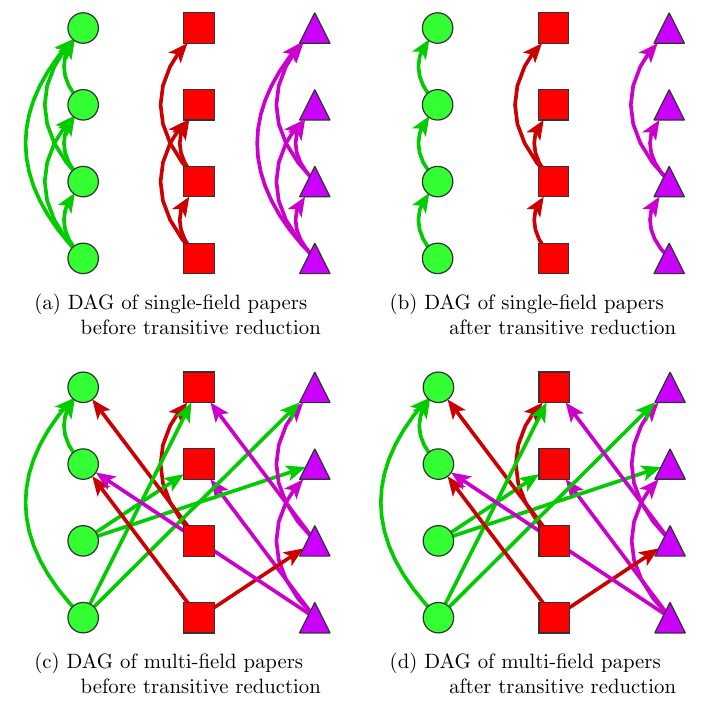}
	\caption{A simple example of how transitive reduction removes more citations between documents in the same field than from multidisciplinary documents. In each of the four citation networks, documents are nodes with older (youngest) at the top (bottom) and documents in the same field are shown in the same colour and with the same symbol in one column. 
    The two citation networks on the left represent two possible examples shown before transitive reduction, while on the right the same citation network is shown but after transitive reduction. 
    The top row is an example where the documents are all single-field documents. Here, transitive reduction removes several edges, reducing the citation count for half the documents. On the bottom row, the nodes now represent multidisciplinary documents, so they often cite documents in another field. No edges are removed under transitive reduction in this case, illustrating the robustness of cross-disciplinary citations under transitive reduction.}
	\label{fig:tr_combined}
\end{figure}

Our proposition is that this difference can be explained by the existence of multidisciplinary documents in these datasets \cite{CGLE14}.
Transitive reduction removes ``inessential edges'', edges that are not required to maintain the connectivity in the citation network. If document $A$ is only of interest to papers from one field, then it is likely many of the papers citing $A$ will also cite each other. So there is a high chance that most of the edges are inessential and will be removed by transitive reduction. If, on the other hand, $A$ is a multidisciplinary paper, then that means it picks up citations from many different fields. It is unlikely that those papers citing $A$ will cite each other as they come from different fields, so this means that the citations to $A$ are much less likely to be inessential. We illustrate this idea in \figref{fig:tr_combined}.

In our approach, we first use clustering to identify documents in the same field and hence to identify multidisciplinary documents. We can then look at collections of documents with a similar citation count, our bins of \eqref{e:Bb}, and measure the change in their in-degree after transitive reduction. Finally, we estimate the mean and standard error of the mean for the change in citation count for each type of document in each bin. A z-score \eqref{e:zbdef} is then used to show if there is a statistically significant difference in the change in the citation count for the multidisciplinary and single field documents.

We first explored the impact of transitive reduction on citation networks by using an artificial model, the PDAG model, described in \secref{s:datamodel} (see also \appref{as:MFCNmodel}) to validate our approach.  We then looked at three datasets: academic papers posted between 1992 and 2003 on the hep-th repository of arXiv \cite{GGK03,KDDcup}, judgements from the U.S.\ Supreme Court (USSC) between 1754 and 2002 \cite{FJ08, FJSJW07} and data on U.S\ patents from the National Bureau of Economic Economic (NBER) from 1970 to 1999 \cite{HJT01}. We first assigned each document to a cluster using a modularity-based community detection method on a bibliographic-coupling network. The cluster labels were then used as proxies for the field of each document. To see if a document was interdisciplinary (cited by papers from many different clusters, many fields) or intradisciplinary (cited by papers from one other cluster, a single field), we used the diversity of the cluster labels held by its neighbours.

Our statistical analysis through the use of z-scores provides clear evidence that intradisciplinary nodes tend to lose more citations than interdisciplinary nodes after transitive reduction, thus supporting our proposition.

The ``reduced in-degree centrality'', that is, the citation count after transitive reduction, has the potential to be a valuable bibliometric metric. Here we have demonstrated that it highlights the interdisciplinarity of documents, an important topic in bibliometrics. This metric could be used to enhance search and recommendation systems by identifying documents that are more accessible to researchers outside their primary field. Furthermore, reduced in-degree centrality could provide valuable insights beyond citation networks. For instance, in ecological food webs, it could help identify keystone species whose extinction would have a significant impact on the ecosystem.

%
%
%



%

%

\subsection*{Data accessibility} 

The data for the three citation networks is available online as cited in the main text and discussed in \appref{as:datasets}: hep-th academic preprints on arXiv \cite{GGK03,KDDcup}, USSC opinions \cite{FJ08} and NBER patents \cite{HJT01}. Copies of the \href{http://doi.org/10.6084/m9.figshare.29663918}{data and code} used in this work are available online \cite{AVE25a}.



%

\subsection*{Funding}

AH is supported by a PhD scholarship from the Saudi Arabian Ministry of Education.
VV would like to acknowledge financial support from the Swiss National Science Foundation and from EPSRC, grant EP-R512540-1.

%
%
%



\begin{center}
	\LARGE\textbf{Appendices}
\end{center}
\appendix
\renewcommand{\thesection}{\Alph{section}}
\renewcommand{\theequation}{\thesection\arabic{equation}}
\renewcommand{\thefigure}{\thesection\arabic{figure}}
\renewcommand{\thetable}{\thesection\arabic{table}}
\numberwithin{equation}{section}
\numberwithin{figure}{section}
\numberwithin{table}{section}
\setcounter{section}{0}
\renewcommand{\theHsection}{\Alph{section}}
\renewcommand{\theHequation}{\thesection\arabic{equation}}
\renewcommand{\theHfigure}{\thesection\arabic{figure}}
\renewcommand{\theHtable}{\thesection\arabic{table}}

\section{PDAG model definition}\label{as:MFCNmodel}


This model is an extension of models and numerical implementations used in \cite{GAE15a,ECV20}.

We start with an initial graph $\Gcal(t=N_\mathrm{init}) = (\Ncal(t),\Ecal(t))$ of $N_\mathrm{init}=120$ nodes but no edges. These nodes are labelled by distinct integers $i$ from $0$ to $(N_\mathrm{init}-1)$, so $n_i\in \Ncal$ is the node with index $i$. Each node $n_i$ also carries a field (cluster) label $c_i$ given by the node's index $i$ modulo the number of fields $N_c=10$, so $c_i = i \mod N_c$. The set of possible field (cluster) labels is therefore $\Ccal = \{ 0,1,2, \ldots , (N_c-1)\}$. 

We also initialise attachment lists, multisets $\Acal_f$ of references to nodes from field $f$. So if $n_i \in \Acal_f$ then $c_i =c_f$. In this case, we initialise these lists with one copy of each node in $\Ncal(t)$ from the appropriate field $f$.

We now add the remaining $(N - N_\mathrm{init})$ new nodes to the graph, one at a time. At each step we create a new graph $\Gcal(t)$ with $(t)$ nodes which starts with a copy of the graph from the previous step, $\Gcal(t-1)=(\Ncal(t-1),\Ecal(t-1))$. We add one  new node $n_{t}$, so with index $t$, to the existing nodes so that $\Ncal(t) = \Ncal(t-1) \cup \{ n_{t}\}$.  The field (cluster) label of the new node, $c_t$ is chosen uniformly at random from the set $\Ccal$ of possible field values. The new node is also classified as being of one of two types; with probability $p_\mathrm{mf} = 1/(N_c+1) \approx 0.0909$ the node is set to be a multi field (multidisciplinary) paper, otherwise it is set to be a single field paper.

Next, we add the edges by creating a bibliography of length $m=12$ for the new node $n_{t}$. The bibliography is made from older papers, target nodes $n_j$ where $j<t$.
If the node $t$ is single-field (SF), then the bibliography is chosen only from older target nodes $j$ (where $j<t$) of the same cluster type $c_t=c_j$, otherwise we have a multi-field paper which can cite any older node. In either case, we choose each older target paper $j$ for the bibliography of node $t$ as follows. 
First, the field $f$ of the cited paper $j$ is chosen to be $f=c_t$ if the new node $t$ is single field, otherwise $f$ is chosen uniformly at random from the set of field labels $\Ccal$. Then, with probability $p=0.45$, the cited paper $j$ is chosen from the last $T_r$ entries in the multi-set $\Acal_f$, otherwise it is chosen from the full set $\Acal_f$. Here, $T_r$ is defined as ``recent papers''. This means that single field papers $n_t$ can only cite papers $n_j$ with the same field $f$, $c_t=c_j$. 

In principle, it is possible for the method to fail at finding $m$ older papers to cite in a reasonable period of time, particularly for single field nodes with a low index where the number of possible older papers it can cite is close to $m$. So in practice, a few earlier nodes have a little short of $m$ documents in their bibliography, but this is almost impossible for nodes $n_t$ where $t \gg m N_c$.

Finally, for every edge added, we update the attachment lists by adding the new node $\mu=12$ times. If the new node $n_t$ is a single-field paper, then it is added $\mu$ times to the attachment list of its own cluster $\Acal_f$ where $f=c_t$. Otherwise, if the new node is a multi-field node, the new node $n_t$ is again added to attachments lists $\mu$ times but this is spread across the attachments lists of different fields, where the fields chosen reflect the fields of the entries in the bibliography\footnote{The field of each paper in the bibliography has a separate entry in a list $\mathtt{B}$, where $\mathtt{B[i]}$ is the field of the $i$-th entry in the bibliography. If the new node is a multifield document, then this list of the fields of bibliometric entries is cycled around $\mu$ times to give a list of $\mu$ fields, that is $f=\mathtt{B[j \mod m]}$ is the $j$-th field chosen with $j$ running over all integers from $0$ to $(\mu-1)$. For each field $f$ chosen, the new node $n_t$ is added to the associated attachment list $\Acal_f$. If $\mu$ is not an integer multiple of $m$, then the new multifield node will not use the fields of the papers in the bibliography equally when choosing which attachment lists to be added to.}. 
Finally, every node $n_j$ in the bibliography of $n_t$ is added once to the attachment list $\Acal_g$ of the old paper's own field, so $g=c_j$.

The purpose of these attachment lists is that, with the exception of the initial nodes, every node appears a total of $(\mu+\kin_j)$ times in these lists. So choosing a node uniformly at random from these lists means we are selecting papers $n_j$ to cite with  a probability proportional to $(\mu+\kin_j)$. This is a form that leads to fat-tailed in-degree distributions for the nodes as shown in simple generalisations of the Price model for bibliometrics \cite{P76,N10}. The point of selecting from the more recent papers is that it is well known that bibliographies tend to have more entries from recent papers than the simple Price model and its variants suggest \cite{GAE15a}. By having this sharp cut-off in time for a fraction $p$ of a bibliography is a crude but simple way to capture this aspect. Given that we add $(m+\mu) = 24$ papers to each attachment list $\Acal_f$ for every new paper $n_i$ in field $f$, 
the recent time scale of $T_r = 52800$ corresponds to  $T_r/(m+\mu)  = 52800/24 = 2200$ new papers in that field.

Note that in some ways, multi-field papers behave in the citation network as if they are in many fields, not just one, even though these papers are also assigned a unique field label $c_i$. This is because they are added to attachment multi-sets $\Acal_f$ that reflect the fields of their bibliographies, not the label $c_i$. The label $c_i$ of a multi-field paper $n_i$ plays a role in the statistics. That is, when we assess the diversity of papers in the artificial network, multi field papers $n_i$ are treated as coming from one field since they carry one label $c_i$. On the other hand, each attachment multi-set $\Acal_f$ contains single-field papers from field $f$ and multi-field papers carrying from any field. This means that a single field paper can cite multi-field papers from other fields. Thus, the out-richness of a single-field paper can be more than one as the bibliography will contain on average about $m .p_\mathrm{mf}(1-1/N_c)$ papers (always multi-field papers) from another field. For the values quoted here, we have $m. p_\mathrm{mf}(1-1/N_c) \approx 12 \times 0.0909 \times 0.9 \approx 0.982$ papers from another field in the bibliography of a single field paper.

\section{Data}\label{as:datasets}

Copies of the \href{http://doi.org/10.6084/m9.figshare.29663918}{data and code} used in this work are available online at
\\
 \url{http://doi.org/10.6084/m9.figshare.29663918} \cite{AVE25a}.

\subsection{Artificial data}

The artificial data is produced by \texttt{python} code implementing the model described in \appref{as:MFCNmodel}.
For the file used in \figref{f:zscoretim}, the parameter values are given in \tabref{at:pdagparam}. These values were chosen to reflect the properties of the hep-th dataset as much as possible and are inspired by the values used in the models of \cite{GAE15a}.
\begin{table}[H]
	\begin{center}
		\begin{tabular}{rc|r}
			\multicolumn{2}{c|}{parameter}      & value \\ \hline
			number of nodes & $N=| \Ncal|$      & 27000 \\
			number of edges & $E=| \Ecal|$      & 322560 \\
			average degree  & $\kav$            & 11.95 \\
			initial nodes   & $N_\mathrm{init}$ & 120 \\
			recent paper time scale
			& $T_r$             & 52800\\
			length of bibliography 
			& $m$               & 12 \\
			fitness         & $\mu$             & 12 \\
			fraction of recent papers in bibliography 
			& $p$               & 0.45 \\
			fraction of multi-field papers 
			& $p_\mathrm{mf}$   & 0.0909 \\
			number fields   & $N_c$             & 10 \\ 
			random seed    &                   & 0\\
		\end{tabular} 
	\end{center}
	\caption[Parameters used for PDAG model]{A list of the parameter values used to generate the example data from the PDAG model. The model is defined in \appref{as:MFCNmodel}.}
	\label{at:pdagparam}
\end{table}



\subsection{Real world data}

The data used for the citation network of academic papers posted on the hep-th section of the arXiv preprint repository between 1992 and 2003 comes from the Cornell KDD Cup website \cite{GGK03,KDDcup}.

The citation network of majority opinions of the United States Supreme Court (USSC) between the years 1754 and 2002 came from links on a 2016 version of a web page \texttt{James H. Fowler's Supreme Court Citation Network Data Page.html}, which cites Fowler and Jeon as authors and asks for citation to two published papers \cite{FJ08, FJSJW07}. 

The dataset used to construct the citation network of patents came from the National Bureau of Economic Research (NBER) \cite{HJT01}, consisting of patents registered in the U.S.\ between 1975 and 1999, with 3,774,768 nodes representing granted U.S.\ patents and 16,518,949 edges representing citations. This also has a single subject category assigned to each patent.

In \tabref{at:dataset_processing_summary} we list some of the basic properties of each network at each stage of our processing.

\begin{table}[ht]
	\centering
	\begin{tabular}{cl|r|r|r|r}
		\textbf{network} & \textbf{stage}     & \textbf{nodes} & \textbf{edges} & \textbf{node\%} & \textbf{edge \%} \\
		\hline
		\href{https://www.kdd.org/kdd-cup/view/kdd-cup-2003/Data}{hep-th}
		\cite{GGK03,KDDcup}
		&  Original                & 27770   & 352807     & 100.0 & 100.0 \\ 
		&  After Edge Correction   & 27770   & 352285     & 100.0 & 99.85 \\ 
		&  After LWCC              & 27400   & 352021 & 98.67 & 99.78   \\ 
		&  After TR                & 27400   & 63247      & 98.67 & 17.93 \\  \hline
		USSC   \cite{FJ08}
		&  Original                & 25417   & 216738     & 100.0 & 100.0 \\ 
		&  After Edge Correction   & 25417   & 216456     & 100.0 & 99.87 \\ 
		&  After LWCC              & 25389   & 216436     & 99.89 & 99.86 \\ 
		&  After TR                & 25389   & 58829      & 99.89 & 27.14 \\  \hline
		NBER   \cite{HJT01}
		&  Original                & 3774768 & 16518949   & 100.0 & 100.0 \\ 
		&  After Edge Correction   & 3774768 & 16518947   & 100.0 & 100.00 \\ 
		&  After LWCC              & 3764117 & 16511740   & 99.72 & 99.96 \\ 
		&  After Degree Filtering  & 1236699 & 8594651    & 32.76 & 52.03 \\ 
		&  After TR                & 1233221 & 5386866    & 32.67 & 32.61 \\ \hline
		PDAG              &  Before TR               & 27000   & 322560     & 100.0 & 100.0 \\ 
		&  After TR                & 27000   & 124855     & 100.0 & 38.71 \\ 
	\end{tabular}
	\caption{Node and edge counts for each dataset at different preprocessing stages. In edge correction, we remove cycles and reverse backward edges to create a DAG. LWCC refers to the extraction of the largest weakly connected component. In degree filtering, we reduce the size of the NBER patent data by removing nodes with degree below ten. Finally we apply transitive reduction (TR).}
	\label{at:dataset_processing_summary}
\end{table}

Edge correction is where we reverse the direction of any edge that cites a document in the future, at least according to the node data we have. We also remove all self-citations. The PDAG model gives a perfect DAG so this correction is not needed. For the real data sets, we only have to remove a very small number of edges, 0.15\% or less.

We always analyse the largest weakly connected component (LWCC), so the next stage is to select this component. Again, the PDAG model gives a single weakly connected component so this is not required. For the other datasets, the change is small, at most a one percent loss, often much less.

The NBER patents dataset is much larger than the other datasets, so we apply a further procedure in this case to extract a smaller relevant network.  Since we only analyse nodes with in-degree greater than nine, we choose to remove nodes and all their associated edges with degree of nine or less from the data.  This is not done recursively, so it is still possible to have nodes with degree less than ten. This leaves us with most of the data to analyse, about one third of the original nodes and half of the edges.

Finally, we apply transitive reduction. As reported in \cite{CGLE14}, for the academic papers in hep-th and the court citations of USSC, we get large drops in the number of edges, as we lose 70\% to 80\% of the edges. The NBER patents however lose far fewer edges, only 37\% of edges were left after the degree filtering. By way of comparison, the artificial PDAG model sits in between, losing 61\% under transitive reduction.


\section{Text-based Clustering}\label{as:textclustering}

One other clustering approach we employ is to use text associated with each document to generate topic labels. The titles of documents, available for all our data, provide too little text to work on. However, the arXiv data set \cite{GGK03,KDDcup} comes with abstracts, and so we group those based on the similarity of their words. 

To do this, we take the words from the abstracts of the hep-th documents, we tokenize and stem the words, excluding all common tokens, those in more than 80\% of documents, and rare tokens, those that are in just 20\% of documents.
We then compute a ``Term Frequency-Inverse Document Frequency'' matrix $T_{\alpha i}$ which is the number of times that token $\alpha$ occurs in document $i$ divided by the logarithm of the number of documents where token $\alpha$ appears at least once. We then use cosine similarity to find the distance $D_{ij}$ between two documents $i$ and $j$
\begin{equation}
	D_{ij}
	=
	1 - \frac{1}{|T_i| \, |T_j|} \sum_{\alpha} T_{\alpha i} T_{\alpha j}
	\, ,
	\quad
	|T_i|^2 = \sum_{\alpha} T_{\alpha i} T_{\alpha i} \, .
\end{equation}
We then used the distance matrix $\Dmat$ to find cluster labels for each document $i$ using the K-Means algorithm\footnote{We tried a variety of other clustering methods, including Density-Based Spatial Clustering of Applications with Noise (DBSCAN) \cite{SSEKX17}, Hierarchical DBSCAN (HDBSCAN) \cite{MH17} and Ordering Points to Identify the Clustering Structure (OPTICS) \cite{ABKS99}. However, we found K-means offered the best trade-off between granularity and cohesion—avoiding both excessive fragmentation (over-clustering) and overly broad groupings (under-clustering).} \cite{S10}. 
The resulting clusters are interpreted as research subfields within the hep-th documents. 
\begin{figure}[H]
	\centering
	\includegraphics[width=\linewidth]{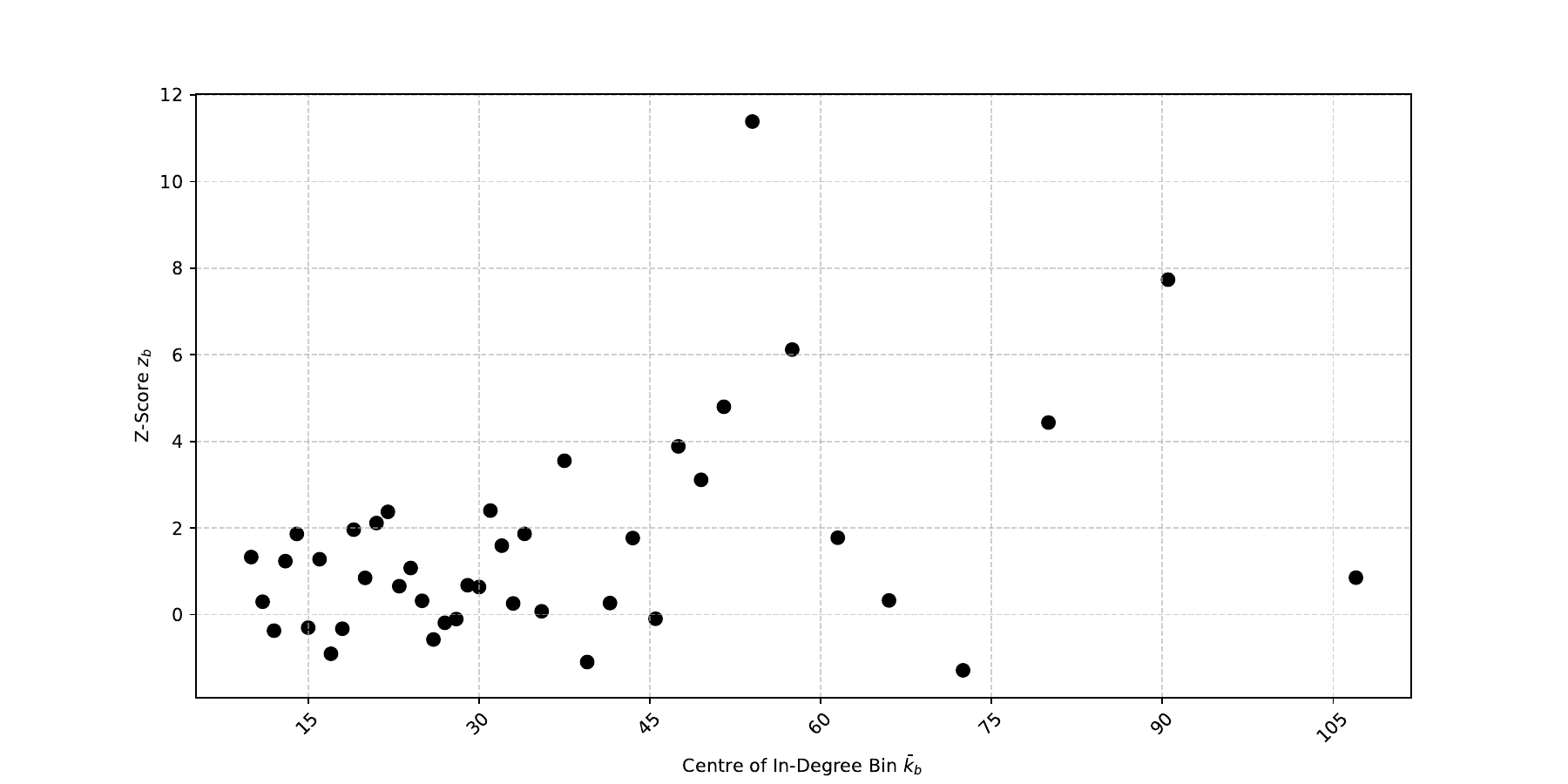}
	\caption{A plot of z-score $z_b$ \eqref{e:zbdef} against bin centre $\kbar_b$ \eqref{e:kbar} for the hep-th DAG using k-means clustering on abstracts with $N_c=7$. 
		The majority of z-score are positive with a few negative points in the centre of the plot.}
	\label{f:zscorehep-thK-means}
\end{figure}
Testing this measure on the hep-th dataset, the results shown in \figref{f:zscorehep-thK-means} suggest a potential trend, though it is not particularly strong. This is likely influenced by limitations inherent to text-based clustering. In our dataset, we observed significant keyword repetition across multiple clusters, which may indicate that text-based clustering is not effectively distinguishing between research subfields. We suspect this issue arises due to the lack of variability in hep-th abstracts, as papers in this domain often revolve around closely related topics. Prior research suggests that abstracts alone may not contain enough distinguishing terms to serve as a reliable basis for clustering, particularly when compared to full-text analysis \cite{AGR05}. This limitation likely contributes to the weaker trend observed in the z-score results.

\end{document}